\RequirePackage{fix-cm}
\documentclass[twocolumn,epjc3]{svjour3}  
\usepackage{cite}% \usepackage{citesort}
\usepackage{graphicx,color,booktabs,math}
\smartqed  % flush right qed marks, e.g. at end of proof
\RequirePackage{graphicx}
\journalname{Eur. Phys. J. C}
\newcommand{\bda}{\begin{\displaymath}
\begin{array}{rl}}
\newcommand{\eda}{\end{array}\end{displaymath}}
\newcommand{\be}{\begin{equation}}
\newcommand{\ee}{\end{equation}}
\newcommand{\bdm}{\begin{displaymath}}
\newcommand{\edm}{\end{displaymath}}
\newcommand{\bea}{\begin{eqnarray}}
\newcommand{\eea}{\end{eqnarray}}
\newcommand{\eq}{\begin{eqnarray}}
\newcommand{\en}{\end{eqnarray}}
 
\newcommand{\fs}{\,.}
\newcommand{\co}{\,,} 
\newcommand{\scs}{\co}

\newcommand{\nnnl}{\nonumber\\[2mm]}

\newcommand{\lttwo}[2]{#1\!<\!#2}
\newcommand{\ltthr}[3]{#1\!<\!#2\!<\!#3}
\newcommand{\gttwo}[2]{#1\!>\!#2}
\RequirePackage[colorlinks,citecolor=blue,urlcolor=blue,linkcolor=blue]{hyperref}

\begin{document}

\title{ Solving integral equations  in $\eta\to 3\pi$%\thanksref{t1}
}


\author{J\"urg Gasser\thanksref{e1,addr1}
        \and
        Akaki Rusetsky\thanksref{e2,addr2} %etc.
}

\thankstext{e1}{e-mail: gasser@itp.unibe.ch}
\thankstext{e2}{e-mail: rusetsky@hiskp.uni-bonn.de}


\institute{Albert Einstein Center for Fundamental Physics,
 Institut f\"ur theoretische Physik, Universit\"at Bern,
 Sidlerstr. 5, CH-3012 Bern, Switzerland \label{addr1}
           \and
 Helmholtz-Institut f\"ur Strahlen- und Kernphysik
 (Theorie) and Bethe Center for Theoretical
 Physics, Universit\"at Bonn, Nussallee 14-16, D-53115
 Bonn, Germany \label{addr2}
%           \emph{Present Address:} if needed\label{addr3}
}

\date{Received: date / Accepted: date}

\maketitle

\begin{abstract}
 A dispersive analysis of $\eta\to 3\pi$ decays
  has been performed in the past by many authors. The
  numerical analysis of the pertinent integral equations
  is hampered by two technical difficulties: i) The
  angular averages of the amplitudes need to be performed
  along a complicated  path in the complex plane. ii) The
  averaged amplitudes develop singularities along the path
  of integration in the dispersive representation of the
  full amplitudes. It is a delicate affair to handle these
  singularities properly, and  independent checks of the
  obtained solutions are demanding and time consuming. In
  the present article, we propose a solution method that
  avoids these difficulties. It is based on a simple
  deformation of the path of integration in the dispersive
  representation (not in the angular average).
Numerical solutions are then obtained rather
straightforwardly. We expect that the  method also works for  $\omega\to 3\pi$.



\end{abstract}

\section{Introduction} \label{sec:Introduction}

The study of the $\eta\to 3\pi$ decay process is
interesting, first and foremost, in the context of the
determination of the quark mass ratio 
\eq\label{eq:Q}
Q^2=\frac{m_s^2-\hat m^2}{m_d^2-m_u^2}\,
,\quad\quad \hat m=\frac{1}{2}\,(m_u+m_d)\, .  \en 
In
order to extract the value of $Q$ to high precision,
it is very important to have a robust control on the
final--state interactions in this decay, which lead to a
strong effect in the width.
 To this end, one would like to have a non--perturbative
framework, allowing the resummation of a certain class of
the final--state interactions to all orders. Dispersion
relations are the ideal tool for this. Incorporating
2--particle unitarity and crossing symmetry then  leads to
a system of coupled integral equations for the 3 isospin
amplitudes in this decay. It took quite some time until
these equations were written down in their final form. The
development started with the pioneering work of Khuri and
Treiman \cite{Khuri:1960zz}. The mathematical structure of
this type of equation was investigated in the following 
decade~\cite{Gribov:1962fu,Bonnevay,Bronzan:1963mby,Kacser:1963zz,Bronzan:1964zz,Aitchison:1964zz,Aitchison:1965zz,Pasquier:1968zz,Pasquier:1969dt}, see also the monograph \cite{Anisovich:2013gha}.
Later, interest in the dispersive method waned. A revival
occurred in the nineties, when  it was  demonstrated 
\cite{Anisovich:1993kn,Kambor:1995yc,Anisovich:1996tx,Walker}
 how dispersion relations allow one to incorporate
 final state interactions of S- and P-waves
 in a reliable and calculable manner. 
  Refs.~\cite{Kambor:1995yc,Anisovich:1996tx} contain a detailed discussion of the role of subtractions in the dispersive representation, while the uniqueness of the solutions is investigated in \cite{Anisovich:1996tx}. 
 Further, current algebra \cite{Osborn:1970nn} and chiral perturbation theory results \cite{Gasser:1984pr} were used to relate the parameter $Q$ to this framework (normalization of the dispersive amplitude), and to get a handle on the subtraction constants.
For a review of the early developments until 1990
 we refer the reader to Ref.~\cite{Anisovich:1993kn},
 see also the lecture notes \cite{Aitchison:2015jxa}. An
 improved representation of the $\pi\pi$ phase shifts
 \cite{Colangelo:2001df,Kaminski:2006qe}, the evaluation
 of electromagnetic corrections \cite{Bissegger:2008ff,Ditsche:2008cq}
 as well as new experimental information on this decay
 \cite{Ambrosino:2008ht,Prakhov:2008ff,Unverzagt:2008ny,Ambrosinod:2010mj,%
Adlarson:2014aks,%
Ablikim:2015cmz,%
Anastasi:2016cdz,%
Prakhov:2018tou}
 triggered new investigations
 of $\eta\to 3\pi$  in the dispersive  framework
 \cite{Lanz_phd,Colangelo:2011zz,Lanz:2013ku,Descotes-Genon:2013uya,Descotes-Genon:2014tla,Guo:2015zqa,Albaladejo:2015bca,Moussallam:2016evb,Guo:2016wsi,Colangelo:2016jmc,Albaladejo:2017hhj}.
  A very comprehensive analysis has recently
 been presented in Ref.~\cite{Colangelo:2018jxw}.
For applications of the dispersive
framework  in other three--body decays, see
Refs.\cite{Niecknig:2012sj,Descotes-Genon:2013uya,Descotes-Genon:2014tla,Niecknig:2015ija,Niecknig:2016fva,Isken:2017dkw,Niecknig:2017ylb}.
Further attempts to incorporate a class of
final state interactions in the $\eta\to
3\pi$  amplitudes may be found	in
Refs. \cite{Roiesnel:1980gd,Gasser:1984pr,Bijnens:2002qy,Bijnens:2007pr,Schneider:2010hs,Schneider:2013phd}.

 The present article is devoted to a discussion of  numerical methods 
 to solve  the integral equations that occur in the formulation
 of Ref.~\cite{Colangelo:2018jxw}.
In this connection, the investigations 
~\cite{Gribov:1962fu,Bonnevay,Bronzan:1963mby,Kacser:1963zz,Bronzan:1964zz,Aitchison:1964zz,Aitchison:1965zz,Pasquier:1968zz,Pasquier:1969dt}
 revealed the following two important aspects:
\begin{enumerate}
\item[i)] In the evaluation
of the angular averages of the amplitudes, the integration
path in $z=\cos\theta$ must be deformed into the complex
plane, in order not to destroy the holomorphic properties
of the amplitudes.  The deformation is fixed by providing
the (eta mass)$^2$ with an infinitesimal positive
imaginary part: $M_\eta^2\to M^2=M_\eta^2+i\delta$~,
with $\delta\to 0^+$ at the end of the
calculation.
\item[ii)]
 Singularities emerge in the angular averaged amplitudes, at
the pseudothreshold $s=(M_\eta-M_\pi)^2$. Some of these
singularities are of a	non--integrable type in the Lebesgue
sense. The (positive) imaginary part of the eta mass,
however, acts as a regulator, that can be removed after
the dispersive	integral has been performed. Handling
these singularities properly is a delicate affair,
as is illustrated  e.g. by the discussions in
Refs.\cite{Kambor:1995yc,Walker,Lanz_phd,Descotes-Genon:2014tla,Schneider:2013phd,Niecknig:2016fva,Albaladejo:2017hhj,Colangelo:2018jxw}.
\end{enumerate}
We shortly mention several previous methods
to obtain solutions of these equations. In
Refs.~\cite{Kambor:1995yc,Walker,Lanz_phd,Colangelo:2011zz,Lanz:2013ku,Niecknig:2012sj,Colangelo:2018jxw},
the equations were solved directly through iterations,
with a careful treatment of the mentioned singularities. In
Ref.~\cite{Descotes-Genon:2014tla,Albaladejo:2017hhj},
integral kernels are introduced, which make the
numerical procedure faster, and which identify the
mentioned singularities at the pseudothreshold
in a clear fashion.  Through the interchange of
the order of integrations, the so--called Pasquier
inversion~\cite{Pasquier:1968zz,Pasquier:1969dt},
it is possible to carry out one integration
 and to obtain integral equations in one
variable that are further solved by iterations (see
Refs.~\cite{Guo:2014mpp,Guo:2014vya,Guo:2015zqa,Guo:2016wsi}
for more details).  Finally, it is possible to
obtain integral equations for the
angular averaged amplitudes instead of the amplitudes
themselves. This technique has recently been invoked in 
Refs.~\cite{Niecknig:2015ija,Niecknig:2016fva}.
Last but not least,  the convergence of
the iterative procedure to solve the equations is not {\it
a priori} guaranteed. The iterative procedure converges
very well (typically, after 3--4 iterations) in the $\eta\to
3\pi$ decays, but the convergence becomes an issue for
decaying particles with heavier masses, or for equations
with more subtractions \cite{Niecknig:2016fva}. For this
reason, e.g., in Ref.~\cite{Niecknig:2015ija}, a matrix
inversion method was used to find a solution beyond the
iterative procedure. These enterprises are numerically
very demanding, in particular for the reasons spelled out above.

In the following, we propose a numerical framework
that avoids the difficulties i) and ii) altogether. 
We show that one may
deform the integration path in the dispersion integral
(not in the angular average) into the complex
plane.
 After choosing a properly deformed path, the integrand becomes
regular, even at $\delta=0$
(except at threshold, where a mild, integrable square
root singularity persists). In addition, the angular
integration can be  carried out in the original
interval $-1 \leq z \leq 1$. Further, discretizing the
integrals through a  Gauss--Legendre quadrature,
the integral equations are transformed into a set of
linear matrix equations, whose solution is easily found
by iterations. The procedure is pretty straightforward,
takes very little CPU time and does not lead to any
spurious irregularities in the solutions\footnote{In the final stage of this work,
we became aware of an article by Aitchison and Pasquier~\cite{Aitchison:1966lpz}, who, more than 50 years ago, noticed the dual feature of deforming integration paths in the angular average and in the dispersive integral. They also noticed that in this manner, the singularity at the pseudothreshold in the dispersive integral can be avoided. See also further  references quoted there. To the best of our knowledge, the method has, however,  never been applied to the problem at hand.}.

\begin{sloppypar}
The idea to avoid singularities in integration through path deformation is not new. An early reference is the work of Hadamard \cite{Hadamard:1898}, whose method was rediscovered 50 years later by investigation of analytic properties of perturbation theory~\cite{Eden:1952,Smatrix:1966}, and then extended and extensively used in S-matrix theory~\cite{Smatrix:1966}.
  The analogue of 
the procedure for the
quantum--mechanical three--bo\-dy problem is well known
since decades~\cite{Hetherington:1965zza,Schmid}.
 As already said, Aitchison and Pasquier mentioned this possibility for Khuri-Treiman-type equations in Ref.~\cite{Aitchison:1966lpz}.
 For the numerical evaluation of loop graphs in Quantum Field Theory,  path deformations in momentum space \cite{Soper:1998ye} as well as in  Feynman parameter space \cite{Nagy:2006xy} are used. 
 All these techniques, including the present one,  may be summarized under the heading ``Applications of Cauchy's integral theorem''.
\end{sloppypar}

The layout of the article is as follows. In
section~\ref{sec:equations}, we display the integral
equations for the $\eta\to 3\pi$ amplitudes in the form
worked out recently in Ref.~\cite{Colangelo:2018jxw}.
We describe in some details the technical
difficulties that one encounters while solving
the equations in section~\ref{sec:hurdles}, while in
section~\ref{sec:avoiding}, we define the deformed path in
the dispersion integral and show that in this manner, the
singularities in the integral equations disappear. 
In section~\ref{sec:solving} we describe the numerical
procedure for solving the equations, whereas a summary and conclusions are 
given in section~\ref{sec:final}. In appendix~\ref{sec:phaseshifts} we comment
 on  the phase shifts used and collect some notation. The holomorphic continuation of 
the integrand is discussed in appendix~\ref{sec:deformation}, path deformations
 in general are investigated in appendix~\ref{sec:choiceofpath}, and the case 
 $\omega\to 3\pi$ is shortly discussed in appendix~\ref{sec:omega}.

\section{The equations}\label{sec:equations} 

\begin{sloppypar}
We start from
the integral equations for the three isospin amplitudes
$M_I(s)$ in the framework specified in \cite{Colangelo:2018jxw},
\end{sloppypar}
\bea\label{eq:M} M_I(s)&=&\Omega_I(s)\left
[P_I(s)+s^{n_I}\int_{4M_{\pi}^2}^\infty
d\mu_I(x)\frac{\hat{M}_I(x)}{x-s-i\epsilon}\right ]\scs
\nonumber\\[2mm]
&&I=0,1,2\fs \eea 
The various quantities are defined as
follows. The Om\-n\`es functions are 
\bea\label{eq:omnes}
\Omega_I(s)=\exp\left(\frac{s}{\pi}\int_{4M_{\pi}^2}^\infty
\frac{dx}{x}\frac{\delta_I(x)}{x-s-i\epsilon}\right),
\eea 
where $\delta_I(x)$ denote the elastic
$\pi\pi$ phase shifts (S-wave for $I=0,2$, P-wave for $I=1$).
The hat--functions  $\hat M_I$ are defined in
terms of angular averages, 
\bea\label{eq:hatM}
\hat{M}_I(x)=\frac{1}{\kappa}\sum_{m,I'}C^m_{II'}(x,\kappa)\langle
z^mM_{I'}\rangle(x),\nonumber\eea 
with
\bea\label{eq:angularaverages} \langle
z^mM_I\rangle(x)&=&\frac{1}{2}\int_{-1}^1
dz z^m M_I(h (x,z))\,;
\nonumber\\[2mm]
 &&I,I',m \in
\{0,1,2\}\scs \eea 
where 
\bea\label{eq:kappa} 
&&h(x,z)=\frac{1}{2}\left[M^2+3M_\pi^2-x+z\,\kappa(M^2,M_\pi^2,x)\right],
\nnnl
&&\kappa(M^2,M_\pi^2,x)=\sqrt{1-\frac{4M_\pi^2}{x}}\sqrt{x-(M-M_\pi)^2}
\nnnl
&&\hspace*{2.15cm}\times\sqrt{x-(M+M_\pi)^2}\fs
\eea 
Finally, the measures
are 
\bea &&d\mu_I(x)=\frac{dx}{\pi
x^{n_I}}\frac{\sin(\delta_I(x))}{|\Omega_I(x)|},\nonumber
\eea 
and
 \bea\label{eq:Mend}
&&\{n_0,n_1,n_2\}=\{2,1,2\}\,;
\nnnl
&& M^2=M_\eta^2 +i\delta\scs\quad
\delta\to 0^+\fs \eea 
The $C^m_{II'}(x,\kappa)$ are
polynomials in $x$ and $\kappa$, and $P_I(s)$ denote
subtraction polynomials, see \cite{Colangelo:2018jxw}. The phase shifts $\delta_I$ are needed as input to solve these equations. We relegate a discussion of them to the appendix~\ref{sec:phaseshifts}, where we also collect some of the notation used below.
\begin{figure}[t]

   \begin{center}
    \includegraphics*[width=7.cm]{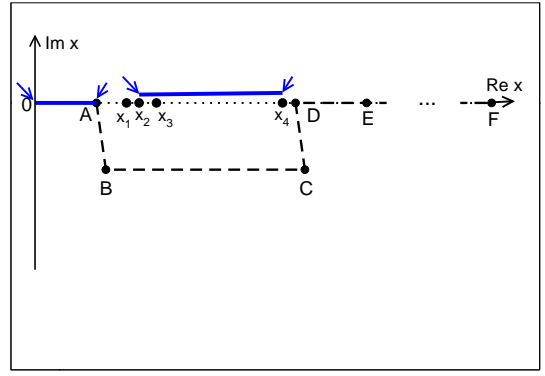}
  \end{center} \caption{Branch points (blue arrows)
  and cuts (solid blue) of the Kacser function
  $\kappa(M^2,M_\pi^2,x)$, and integration paths. We
  connect the branch points with cuts in the intervals
  $[0,4M_\pi^2]$ and $[(M-M_\pi)^2,(M+M_\pi)^2)]$. The
  original path of integration for the dispersive
  representation (\ref{eq:M}) runs along the real line,
  from A to DEF (dotted black), the deformed path along  ABCDEF (dashed black). See appendix~\ref{sec:phaseshifts} for $x_i$ and $A,B,\ldots F$.
The figure is not drawn on scale.} \label{fig:1}
 \end{figure}

\begin{figure}[t]
  \begin{center}
  ~ \hspace*{-.7cm} \includegraphics*[width=8.cm]{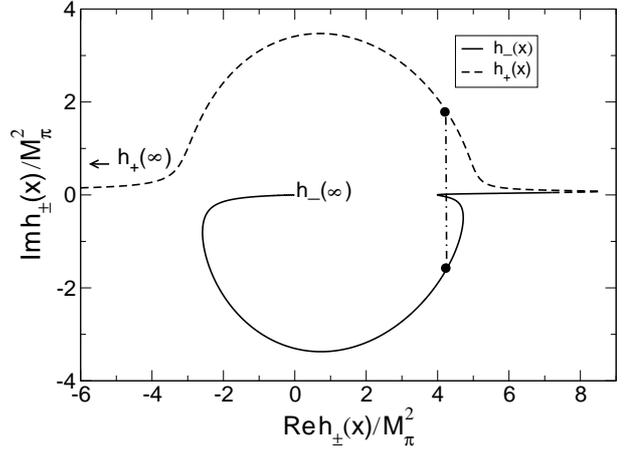}
  \end{center} \caption{Real and imaginary parts of
  the endpoints $h_\pm(x)$ in (\ref{eq:endpoints})
  in the angular
 averages (\ref{eq:angularaverages}), for $x\in
 [4M_\pi^2,\infty)$. For better visibility, we have
 kept in the eta mass a positive imaginary part. Solid
 (dashed) line: lower (upper) endpoint. The dash--dotted
 line connects $h_-(x)$ with $h_+(x)$, with $\ltthr{x_2}{x}{x_3}$. 
It is seen that this line crosses the original
 path of integration in the dispersive representation
 (\ref{eq:M}). The same happens for $\ltthr{x_1}{x}{x_2}$. See
 text for details.}
\label{fig:2} \end{figure}

\begin{figure}[t]
  \begin{center}
    \includegraphics*[width=7.cm]{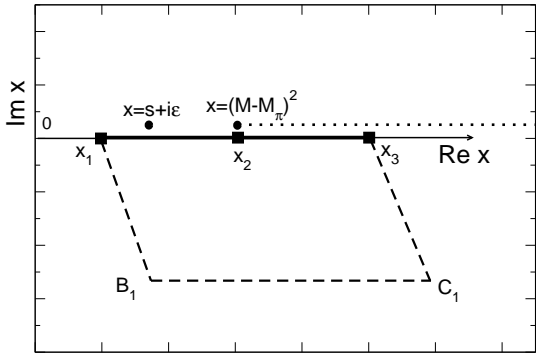}
  \end{center} \caption{Integration paths for the singular
  integral $G(s)$ in (\ref{eq:G(s)}). The original
  path runs along the real line, from $x_1$ to $x_3$
  (solid line). 
The upper horizontal line (dotted) stands for the cut attached to
  the branch point  at $x=(M-M_\pi)^2$.
The two singularities of the integrand are
  indicated  with filled circles. Since these are located
  in the upper complex plane, the path may be deformed
  into the lower half plane, e.g., into the polygonal line $x_1B_1C_1x_3$ (dashed line), without changing the value of the integral. See appendix \ref{sec:phaseshifts}
 for $x_i,B_1,C_1$.
 \label{fig:3} }\end{figure}

\begin{figure}[t]
  \begin{center}
    \includegraphics*[width=7.cm]{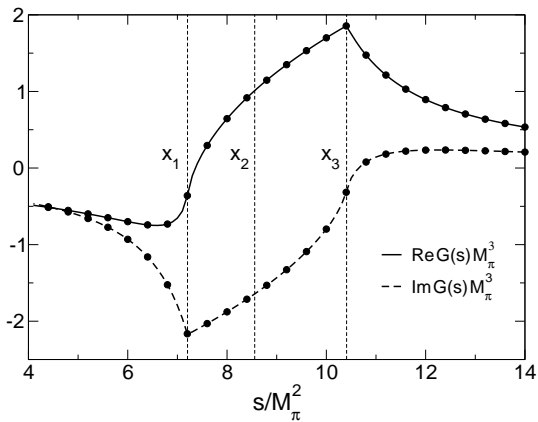}
  \end{center} \caption{Analytical and numerical results for the function $G(s)$ 
in Eq.~\ref{eq:G(s)}. The solid and dashed lines denote the real and imaginary part of the analytical result \cite{Colangelo:2018jxw}, the filled circles denote the result from the Gauss-Legendre integration along the polygonal line $x_1B_1C_1x_3$ in Fig.(\ref{fig:3}).
\label{fig:3b}}
\end{figure}

\begin{figure}[t] 
  \begin{center}
  ~\hspace*{-0.8cm}  \includegraphics*[width=7.6cm]{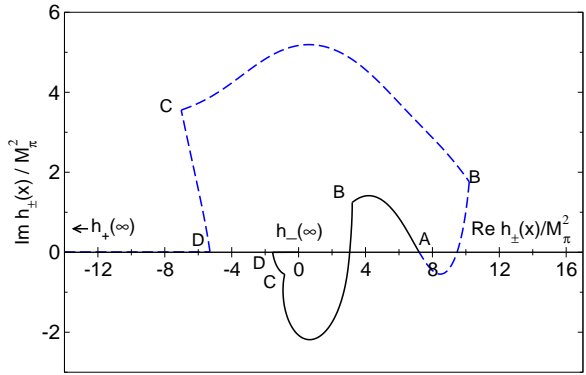}
  \end{center} \caption{Real and imaginary part of the
  endpoints $h_\pm(x)$ in (\ref{eq:endpoints}), if the
  integration in the dispersive integral (\ref{eq:M})
  is performed along the deformed path ABCDEF in
  Fig.~\ref{fig:1}. Compare with Fig.~\ref{fig:2},
  which shows the situation in the standard case.}
  \label{fig:4} \end{figure}

\begin{figure}[t]
  \begin{center}
    \includegraphics*[width=7.cm]{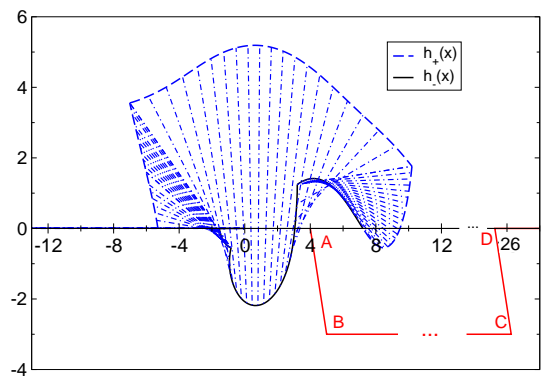}
  \end{center} \caption{Displayed are some
  straight lines (dash--dotted) that connect the
  lower and upper ends $h_\pm(x)$ in the angular
  averages (\ref{eq:angularaverages}), see also
  Fig.~\ref{fig:4}. Axes in units of $M_\pi^2$. The eta
  mass is real. It is seen that the dash--dotted lines do
  not cross the path ABCD and hence they stay within the
  holomorphy domain of the amplitudes $M_I(s)$. One can
  therefore perform the angular averages in their original
  form.} \label{fig:5}
\end{figure}

\begin{figure}[t]
  \begin{center}
    \includegraphics*[width=7.cm]{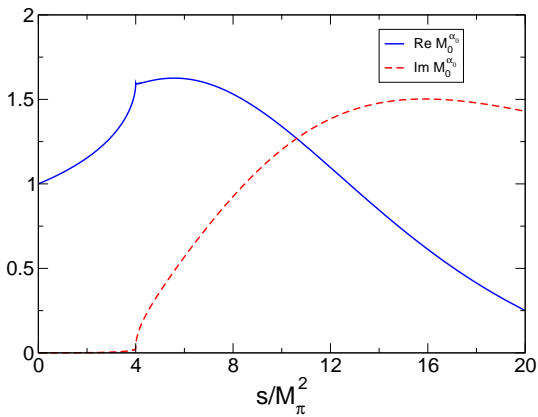}
  \end{center} \caption{Solution of the isospin
  zero amplitude. Notation
  as in Ref.~\cite{Colangelo:2018jxw}. } \label{fig:M0}
  \end{figure}

\section{Technical hurdles}\label{sec:hurdles} During
the numerical, iterative solution of the equations
(\ref{eq:M})-(\ref{eq:Mend}), two main	problems occur.

i) The angular averages $\langle z^m M_I \rangle(x)$
amount to an integration along the path
 $ h(x,z) $ of the argument of the amplitudes $M_I$. For
 fixed $x$, the argument runs along a straight line,
 from $h_-(x)$ to $h_+(x)$, where
\bea\label{eq:endpoints}
&&h_\pm(x)=\frac{1}{2}\left[M^2+3M_\pi^2-x\mp\kappa(M^2,M_\pi^2,x)\right]\fs
\eea 
The Kacser function $\kappa(M^2,M_\pi^2,x)$
\cite{Kacser:1963zz} is holomorphic in the complex
$x$--plane, cut along the real axis from $x=0$ to $x=4
M_\pi^2$ and along a straight line from $x=(M-M_\pi)^2$
to $x=(M+M_\pi)^2$ , see Fig.~\ref{fig:1}\footnote{The branch points at $x=(M\pm M_\pi)^2$ are located in the upper half plane because of the prescription $M_\eta^2\to M^2=M_\eta^2+i\delta,~\delta>0$. It is here that the sign of the imaginary part in $M^2$ matters.}.
 Therefore,
fixing its value at some point in the complex $x$--plane
renders it unique in the cut plane. In the following, we
use the convention 
 \bea \kappa(M^2,M_\pi^2,x)=x + O(1)\co
x \to \infty\fs \eea 
In Fig.~\ref{fig:2}, we display the
endpoints $h_\pm(x)$ as the integration variable $x$
runs from the threshold to infinity. Because
$\kappa(M^2,M_\pi^2,x)$ is holomorphic in the cut $x$
plane, the endpoints $h_\pm$ move -- for $\delta > 0$ --
along curves that are infinitely often differentiable
with respect to the real variable $x$. The dash--dotted
vertical line connects the endpoints $h_\pm(x)$ with
$\ltthr{x_2}{x}{x_3}$. It is seen that a straight line between
$h_-(x)$ and $h_+(x)$ crosses the integration path in the
dispersive representation (\ref{eq:M}). This
is also the case for $\ltthr{x_1}{x}{x_2}$.  The
problem is discussed at many places in the literature -- early references are \cite{Gribov:1962fu,Bronzan:1963mby}.
 Its solution amounts to deform the path in the $z$
integration (or in the integration over the variable $h$
after a change of  variables $z\to h$), such that this
crossing is avoided. We do not discuss this point any
further here.

ii) The second problem arises after the angular integration
has been performed in one way or the other. The angular
averages develop singularities of the type 
\bea 
\hat{M_I}(x) \sim \frac{K}{[(M - M_\pi)-x)]^{p}
}\scs
\eea 
where $p=(\frac{1}{2},\frac{3}{2},\frac{1}{2})$ for $I=(0,1,2)$,
and $K$ denotes
 a constant.
  These singularities  then show up in the
 dispersive integrals (\ref{eq:M}).
It turns out that,
 after the integration over  $x$ has been performed,  the
 limit $M \to M_\eta$ exists. [See e.g. Refs.~\cite{Kambor:1995yc,Colangelo:2018jxw}, where the dispersive integral has been worked out explicitly.] We
 note that the case $p=\frac{3}{2}$ corresponds to a
 non--integrable singularity at $\delta=0$: one is not
 allowed to interchange the limit $\delta\to 0^+$ with
 the integration over $x$. These singularities render the
 standard procedure to construct a numerical  solution of
 the integral equations rather delicate and cumbersome.
 In the following, we present a method to solve these
 equations   in an easy and straightforward manner, that
 avoids the problems i) and ii) altogether. [Including higher partial waves leads to even stronger singularities~\cite{kubis_private}.  
Our method also covers these cases, see the following section.]

\section{Avoiding singular integrals}\label{sec:avoiding}

There is a simple way to cope with the singularities
of the angular averages. To illustrate, we consider
the integral \cite{Kambor:1995yc,Colangelo:2018jxw} 
\bea\label{eq:G(s)}
&&G(s)=\int_{x_1}^{x_3}\frac{g(x)\, dx}{(x-s-i\epsilon)((M-M_\pi)^2-x)^{3/2}}\scs\nnnl
&&g(x)=\frac{(x-x_1)(x_3-x)}{(x_2-x_1)(x_3-x_2)}\fs
\eea 
We display in Fig.~\ref{fig:3} the singularities of
the integrand with two black dots in the complex $x$-plane:
one at $x=s+i\epsilon$, the second at 
$x=(M-M_{\pi})^2\stackrel{\delta\to 0^+}{\rightarrow}x_2$. We
have also drawn a cut that emerges from the branch
point at this second singularity, reaching out to $x =
\infty+i\delta$ (dotted line).  The path
 of integration is indicated with a solid line, from $x_1$
 to $x_3$. We now observe that the integrand is holomorphic
 in the complex half--plane Im$(x)\!<\!\epsilon,\delta$.
 Therefore, we may deform the path of integration into the polygonal line  $x_1B_1C_1X_3$ (dashed), without
 changing the value of the integral. There are then no
 singularities anymore on the integration path, and we may
 set $\epsilon=\delta=0$ before performing 
the integration.
This proves
 that the limit $\epsilon,\delta\to 0^+$ of the original integral
 (\ref{eq:G(s)}) exists \cite{Kambor:1995yc,Colangelo:2018jxw}. 
[It is obvious that this remains true if 
the exponent $3/2$ in (\ref{eq:G(s)}) is replaced by any  $p\in \C$.] 
 On the other hand, approaching the real axis from below, one encounters a pinch singularity at $s=(M-M_\pi)^2$, which results in a singular behaviour of the function $G(s)$.

 A numerical
 integration along the dashed path does not pose any
 problems, because the integrand is smooth as a function
 of $x$. To illustrate, we display in Fig.~\ref{fig:3b} the real part (solid line)
 and imaginary part (dashed line) of the function $G(s)$ (analytic expression 
is given in \cite{Colangelo:2018jxw}), in units of the pion mass. For comparison, 
the result of the numerical integration along the dashed polygonal 
line with 160 Gauss-Legendre points is shown with filled circles. It is seen that the agreement is perfect.

We now use the same method in the original  equations 
(\ref{eq:M}). Suppose that the integrand in (\ref{eq:M}) is
 holomorphic in some region Im($x)\!<\! 0$ (we discuss this point in 
appendix~\ref{sec:deformation}). Then we can avoid the singularities generated
by the zeros in the function $\kappa(M^2,M_\pi^2,x)$ by
deforming the path as shown in Fig~\ref{fig:1}. Aside
from avoiding the singularities in $\hat {M}_I(x)$,
this procedure has the following advantage.  Consider the
endpoints of the angular integration as they now occur on
the deformed path, see Fig.~\ref{fig:4}. There, we display
the two endpoints $h_\pm(x)$. It is seen that the problem
with leaving the holomorphy domain of $M_I(s)$ does not
occur anymore. See also Fig.~\ref{fig:5}, where we display
some of the paths $h(x,z),~ -1\leq z \leq 1$, that connect
$h_\pm(x)$. We conclude that one  can solve the	equations
(\ref{eq:M}) in their original form, avoiding complicated
path deformations and  singular integrals, provided
that the path of integration in (\ref{eq:M}) is deformed
properly -- e.g. according to Fig.~\ref{fig:1}.

 The deformed path displayed in Fig.~\ref{fig:1} is obviously  
not the only one with these properties  -- see appendix~\ref{sec:choiceofpath} for a discussion of this point.

\section{Solving the integral equations}\label{sec:solving}
Finally, we make use of the fact that now, the integrands
are smooth along the path of integration for $s<D$, except at the
threshold $s=4 M_\pi^2$, where an integrable singularity of the type
$\mbox{const}/\sqrt{x-4M_\pi ^2}$ occurs (note that $\hat M_I(x)\to $ const. as $x\to 4M_\pi^2)$. The singularity can be tamed with a variable transformation $z=A+(B-A)\tau^2\scs 0\le\tau\le 1$.  Therefore,
an integration using the Gauss--Legendre method \cite{NAG}
is adequate.
 As we now show,
 this method has the further advantage that
the integral equations boil down to a matrix equation,
with matrix elements that need to be evaluated only once,
before the iteration, which then becomes trivial.

We use Gauss--Legendre quadrature both for integration
over $x$ and over $z$. The integration path over $x$ is
split into linear pieces: $AB$, $BC$, $CD$ and $DF$ (the
point $F$ corresponds to the upper limit of integration:
the phase shifts $\delta_I(x)$ are equal either to $0$
or to $\pi$, if $x$ moves right to $F$).
 The variable $x$ from each interval is mapped to the
 interval $[0,1]$, and
the Gauss--Legendre mesh points $x_i$ and weights $w^x_i$
are used to carry out the integration.	The number of
points on these intervals is $N_{AB}$, $N_{BC}$, $N_{CD}$
and $N_{DF}$, respectively. Further, the integral over $z$
 for a given $x$ always runs from $-1$ to $1$.  The mesh
points and weights are denoted by $z_a$ and $w^z_a$,
respectively, and the number of mesh points is chosen
to be  $L_{AB}$, $L_{BC}$, $L_{CD}$ and $L_{DF}$ for $x$
belonging to the one of the above intervals.

It is useful to introduce a multi--index, 
\eq
\alpha=(i,a)\, , \quad h_\alpha=h(x_i,z_a)\,
,\quad w_\alpha=\frac{1}{2}\,w^x_iw^z_a\, .
\en Here, \eq \alpha&=&1,\cdots N\, ,
\nnnl
N&=&N_{AB}L_{AB}+N_{BC}L_{BC}+N_{CD}L_{CD} +N_{DF}L_{DF}\,
.  
\nonumber\\
&&\en 
Combining the equations (\ref{eq:M}) and
(\ref{eq:hatM}), we may write 
\eq\label{eq:s_alpha}
M_I(s)=\Omega_I(s)\biggl[P_I(s)+\sum_{\alpha,I'}w_\alpha
K_{II'}(s,h_\alpha) M_{I'}^\alpha\biggr]\, , \en
 where
\eq K_{II'}(s,h_\alpha)&=&\sum_m \frac{s^{n_I}}{\pi\kappa(M_\eta^2,M_\pi^2,x_i)
x_i^{n_I}}\,C^m_{II'}(x_i,z_a)
\nnnl
  &\times&\frac{\sin(\delta_I(x_i))}{|\Omega_I(x_i)|}\frac{z_a^m}{x_i-s}\,
  .  \en 
\begin{sloppypar}
Finally, letting the variable $s$ run
  over the set $h_\alpha,~\alpha=1,\ldots
  N$ and introducing the notations
  $R_{II'}^{\beta\alpha}=w_\alpha\Omega_I(h_\beta)K_{II'}(h_\beta,h_\alpha)$
  and $G_I^\beta=\Omega_I(h_\beta)P_I(h_\beta)$, we can
  write down Eq.~(\ref{eq:s_alpha}) as a matrix equation
\end{sloppypar}
  \eq\label{eq:matrix} M_I^\beta=G_I^\beta+\sum_{\alpha,I'}
  R_{II'}^{\beta\alpha}M_{I'}^\alpha\, ; \,\quad\quad
  \beta=1,\ldots,N\fs \en 
This equation generates iteration
  series for the vectors $M_I^\beta$, $I=0,1,2$, which
  are rapidly convergent for $\eta\to 3\pi$ (note that
  the quantities $R_{II'}^{\beta\alpha}$ need to
  be evaluated only once for a given set of phase
  shifts).  
 For $s\not\in [D,F]$, the amplitudes 
   can then be directly 
  constructed from $M_I^\beta$ by using the dispersive
  representation (\ref{eq:M}).
For $s\in[D,F]$, the angular averaged amplitude must be interpolated 
to perform the Cauchy integral. Moreover, if one manages to invert the
  large matrix $R_{II'}^{\beta\alpha}$ numerically,
  Eq.~(\ref{eq:matrix}) can be solved directly, without
  iterations, even if the latter do not converge~\cite{Niecknig:2015ija}. 
Because
  there was no need to do so in our case, we sticked to
  the iteration procedure.

Finally in Fig.~\ref{fig:M0}, we display the isospin zero
amplitude for one particular choice of the subtraction
polynomials, and for a specific choice of the $\pi\pi$
phase shifts. The amplitude agrees with
the one constructed by Lanz \cite{Lanz}, except near the
threshold, where the cusp structure is different,  and
near the pseudo--threshold in the $I=1$ channel.

\section{Summary and conclusions}\label{sec:final} 
\begin{enumerate}
\item
We have
considered numerical aspects of the integral equations
Eqs.(\ref{eq:M})--(\ref{eq:Mend}) that govern $\eta\to 3\pi$
decays \cite{Colangelo:2018jxw}. The standard approach
to solve these equations numerically  is confronted with two main
technical hurdles: angular averages along complicated paths
in the complex plane, and singularities of the integrand
near the integration path in the dispersive representation.

\item
Holomorphicity of the phase shifts in the low
energy region allows one to deform the original path of
integration in the dispersive representation (not in the
angular averages). Both problems disappear~\cite{Aitchison:1966lpz}: The angular
averages can be performed in their original form, and
there are no nearby non--integrable singularities on the integration path.

\item
As the integrands are smooth, a Gauss--Legendre
integration becomes feasible. The integral equations turn
into a matrix equation, whose iterative solution can be
obtained very efficiently.

\item
We have constructed the 6 fundamental
solutions \cite{Colangelo:2018jxw} of
Eqs.(\ref{eq:M})--(\ref{eq:Mend}) for several sets of phase
shifts, in the region $\ltthr{0}{s}{20 \, M_\pi^2}$. The results generally agree
with the solutions obtained by Lanz \cite{Lanz}.

\item
Tables with the fundamental solutions for 8 different sets of phase shifts  
are submitted as ancillary files, together with tables for the phase shifts used.

\item
As we show in appendix~\ref{sec:omega}, the very same technique is expected to also  work for $\omega \to 3 \pi$ \cite{Niecknig:2012sj}. It remains to be seen to what extent it can be applied to other three--body decays and to the calculation  of form factors~\cite{Niecknig:2012sj,Descotes-Genon:2013uya,Descotes-Genon:2014tla,Niecknig:2015ija,Niecknig:2016fva,Isken:2017dkw,Niecknig:2017ylb}.

\end{enumerate}

\section*{Acknowledgments} The authors	thank Gilberto
Colangelo, S\'ebastien Des\-co\-tes-Genon, Christoph Hanhart, Bastian Kubis, Stefan Lanz, 
Heinrich Leutwyler, Ulf--G. Mei{\ss}ner,  Malwin Niehus  and Zoltan
Kunszt for 
discussions and/or useful comments  on the manuscript and  for information on the $\pi\pi$ phase shifts. 
 We thank Stefan Lanz for
providing us with his Fortran code to construct
the numerical solutions using the standard approach, 
and for  data files with his  fundamental solutions. 
We appreciate the support from Emilie Passemar to handle the hyperreferences.
 Bachir Moussallam has pointed out to us that there are spikes in our former fundamental solutions.
 A.R. acknowledges the support from the DFG
(CRC 110 ``Symmetries and the Emergence of Structure in
QCD''), from  Volkswagenstiftung under contract no. 93562
and from Shota Rustaveli National Science Foundation
(SRNSF), grant no. DI-2016-26.
 J.G.   thanks the
HISKP at the University of Bonn for warm hospitality. Part
of this work was performed during his stay there.

\begin{appendix}

%before appendix

\newcounter{zahler}
\renewcommand{\thesection}{\Alph{zahler}}
\renewcommand{\theequation}{\Alph{zahler}.\arabic{equation}}
\renewcommand{\thefigure}{\Alph{zahler}.\arabic{figure}}
\setcounter{zahler}{0}

%before each section in appendix

\setcounter{equation}{0}
\setcounter{section}{0}
\setcounter{figure}{0}
\addtocounter{zahler}{1}

\section{Notation and phase shifts}\label{sec:phaseshifts}
In the text and in figures, we use the symbols
\bea\label{eq:xi}
&&x_1=\frac{1}{2}(M_\eta^2-M_\pi^2)\scs\quad\quad 
x_2=(M_\eta-M_\pi)^2\scs\nnnl
&&x_3=M_\eta^2-5M_\pi^2\scs\quad\quad
x_4=(M_\eta+M_\pi)^2\fs
\eea
The new integration path is fixed by the vertices
\bea \label{eq:abcd1}
&&A=4M_\pi^2\scs \quad\quad
B=A+M_\pi^2(1-3i)\scs
\nnnl
&&C=D+M_\pi^2(1-3i)\scs\quad\quad
D=x_4+M_\pi^2\scs 
\nnnl
&&E=(.8 \mbox{GeV})^2\scs\quad\quad
F=(1.7 \mbox{GeV})^2\fs
\eea
In section \ref{sec:avoiding}, we set
\bea
B_1=x_1+M_\pi^2(1-3i)\scs\quad
C_1=x_3+M_\pi^2(1-3i)\fs
\eea 
In the case $\omega\to 3\pi$, we use
\bea
&&y_i=x_{i|M_\eta^2\to M_\omega^2}\scs\nnnl
&&A_2=4M_\pi^2\scs\quad\quad
B_2=A_2+M_\pi^2(1-7i)\scs\nnnl
&& C_2=D_2-7iM_\pi^2\scs\quad\quad
D_2=y_3+2.5 M_\pi^2\fs
\eea
The elastic S-wave (P-wave) $\pi\pi$ phase shifts $\delta_{0,2}$ ($\delta_1$) are needed as an input to solve (\ref{eq:M}) -- (\ref{eq:Mend}). We rely on the  phase shifts used in Ref.~\cite{Colangelo:2018jxw}: In the low energy region 
 $\lttwo{x}{E} $,  we use a Schenk parameterization\cite{Schenk:1991xe}. Above $x=F$, the phase shifts are set to a constant,
\bea
 \delta_{0,1}=\pi\scs\quad\quad
\delta_2=0\,\,;\quad x\geq F\fs
\eea
For $E \leq x \leq F$, we use phase shifts from \cite{Colangelo:2018jxw}.
Tables with these phase shifts at discrete values of energies
between E and F, as well as the Schenk parameters to describe
the phase shifts below E are provided as
ancillary files, together with the pertinent fundamental solutions.

\setcounter{equation}{0}
\setcounter{section}{0}
\setcounter{figure}{0}
\addtocounter{zahler}{1}

\renewcommand{\thesection}{\Alph{zahler}}
\renewcommand{\theequation}{\Alph{zahler}.\arabic{equation}}
\renewcommand{\thefigure}{\Alph{zahler}.\arabic{figure}}

\section{Deforming the path of
integration}\label{sec:deformation}

\begin{figure}[t]
  \begin{center}
~\hspace*{-0.7cm}    \includegraphics*[width=8.cm]{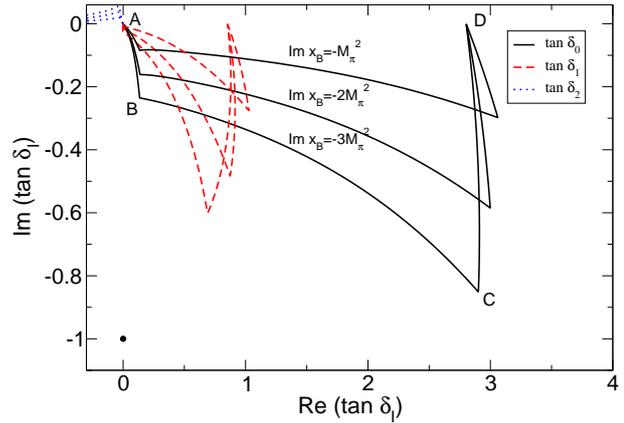}
  \end{center} \caption{Real and imaginary parts
  of $\tan \delta_I$ along the path ABCD, see
  Fig.~\ref{fig:1}. $x_B$ denote the imaginary part along
  the horizontal line BC. The black dot in the lower left
  corner denotes the position of the singularity  $\tan
  \delta_I=-i$ (the $f_0(500)$ 
in the case of $I=0$ \cite{Caprini:2005zr,Pelaez:2015qba}). It is seen 
that the singularity is never
  hit in the region spanned by the path ABCD with the choice (\ref{eq:abcd1}) for the 4 vertices. Schenk
  parameterization \cite{Schenk:1991xe}
is used, with parameters from \cite{Colangelo:2001df}, see appendix~\ref{sec:phaseshifts}.}
 \label{fig:6}
  \end{figure}

Here, we wish to
show that the original path of integration in
 (\ref{eq:M}) can be deformed into the lower complex
 $x$--plane. For this to achieve, we must prove that the
 integrand is holomorphic there.  As an input, we need the
 phase shifts $\delta_I(x)$ and a starting value of the
 hat--functions $\hat M_I(x)$. For the latter, we use
 the current algebra expressions, which are polynomials
 in $x$ and thus even entire functions. Concerning the
 phase shifts, see appendix~\ref{sec:phaseshifts}.
 The Schenk parameterization used below 800 MeV 
has the virtue that the corresponding expression for the
phase shift is holomorphic in a strip of  the 
complex $x$--plane, cut for $x<4M_\pi^2$ and $x>700$ MeV. This is so
because the Schenk parameterization provides a holomorphic
expression for the tangent of the phase shifts. The phase
shift itself is then obtained via 
\bea \delta_I(x)=\frac{1}{2i}\log{\frac{1+i
\tan \delta_I(x)}{1-i \tan \delta_I(x)}}\fs \eea 
As
long as $\tan \delta_I(x)\neq \pm i$, $\delta_I(x)$ is
holomorphic. We display in Fig.~\ref{fig:6} the complex
quantities $\tan \delta_I(x)$ along the polygonal path
ABCD displayed in Fig.~\ref{fig:1}, and along similar
paths which are nearer to the real line. It is seen that
one does not hit any singularity. We have checked
that the parameterization in terms of conformal variables
as worked out in Ref.~\cite{GarciaMartin:2011cn} leads to
the same conclusion.

We conclude that  $\sin{\delta_I}$ is holomorphic in
that region as well. Consider now the denominator of
the measure, $|\Omega_I(x)|$. At first, it sounds
surprising that one can continue analytically the
modulus of a complex function. The point is that this
quantity stands for 
\bea |\Omega_I(x)|=\exp\left
( \frac{x}{\pi}-\hspace{-4.mm}\int_{4M_\pi^2}^\infty\frac{dy}{y} \frac{\delta_I(y)}{y-x}\right
), \eea\label{eq:pvintegral}
 where
$-\hspace{-3.2mm}\int$ denotes a principal value
integral. The relevant quantity is thus the integral
\bea G_I(x)=\frac{x}{\pi}-\hspace{-4.mm}\int_{4
M_\pi^2}^\infty\frac{dy}{y}\frac{\delta_I(y)}{y-x}\fs
\eea
 It is useful to first consider the related
ordinary integral 
\bea H_I(x)=\frac{x}{\pi}\int_{4M_\pi^2}^\infty
\frac{dy}{y}\frac{\delta_I(y)}{y-x}\scs\quad\quad
 x\in\C\fs
\eea 
The function $H_I(x)$ is holomorphic
in the complex $x$--plane, cut along the real
axis for $x \geq 4M_\pi^2$. On the upper rim
of the cut, $G_I$ and $H_I$ are closely related:
\bea\label{eq:HP} H_I(x+i0^+)=i\delta_I(x)+G_I(x)\scs\quad
x\in[4M_\pi^2,\infty)\fs \eea 
Consider now the interval
$[A,D]$ in Fig.~\ref{fig:1}. We can continue here $H_I(x)$
across the cut, because $\delta_I(x)$ are holomorphic there,
and the integration path $\int_A^D$ can be deformed into
the lower half plane, e.g., into the polygonal line ABCD
in Fig.~\ref{fig:1}. The relation (\ref{eq:HP}) shows
that $G_I$ and thus $|\Omega(x)|$ can holomorphically be
 continued as well. The continuation is unique.

\begin{sloppypar}
In practice, the continuation can be performed
rather easily: One adds and subtracts the phase shift
$\delta_I(x)$ in the integrand of $G_I(x)$ and obtains
\end{sloppypar} 
\bea
G_I(x)&=&-\frac{\delta_I(x)}{\pi}\ln{\frac{x-4M_\pi^2}{4M_\pi^2}}
\nnnl
&+&\frac{x}{\pi}\int_{4M_\pi^2}^\infty\frac{dy}{y}\frac{1}{y-x}[\delta_I(y)-\delta_I(x)]\fs
\eea
 This identity holds  for  $x\in [4M_{\pi}^2
,\infty)$. In the integral on the right hand side, one need
not use the principal value prescription. The right--hand
side is holomorphic in the region where the phase shift
is holomorphic.

  Finally, the Omn\`es functions $\Omega_I(x)$ which
  enters  the iterations in the region covered by the
  dashed-dotted blue lines in Fig.~\ref{fig:5} may be evaluated in
  a manner analogous to $|\Omega_I(x)|$ just discussed:
  Below the threshold, use the integral representation
  (\ref{eq:omnes}), as well as above threshold, for
  Im$(x)> 0$. In the case
of Im$\lttwo{(x)}{0}$, Re$\gttwo{(x)}{4M_\pi^2}$, proceed as in the case
of the principal value integral just discussed.

We conclude that we may indeed	deform the path of
integration as is indicated in the Fig.~\ref{fig:1}. [In
order to render the proof watertight in view of
the square--root singularities at the threshold
 $x=4M_\pi^2$, one may slightly change the paths. Let
$A'=4M_\pi^2+\epsilon,~ \epsilon>0$. Then  ABCDF$\to$ AA$'$BCDF. Numerically, this
change is irrelevant, and we stick to $\epsilon=0$.]

We have checked that the results do not change under a
change of the polygonal line ABCD.

The following two remarks are in order at this place.
\begin{enumerate} \item We do {\it {not}} assume that
the phase shifts are holomorphic at all energies. As
mentioned, we glue the phase shifts continuously together, using
different parameterizations. Only in the low--energy
region	do we rely on a holomorphic parameterization.
\item One might argue that we extrapolate data from the
real line to the polygonal line ABCD, which would be an
unstable procedure. Instead, we make  use of a holomorphic
parameterization to render the integrations easier to
perform. For a given parameterization, the result is
the same, whether performed in the standard manner, or
using a deformed path in the dispersion relation. The
latter method is, however,  by far more efficient.  \end{enumerate}

\setcounter{equation}{0}
\setcounter{section}{0}
\setcounter{figure}{0}
\addtocounter{zahler}{1}

\renewcommand{\thesection}{\Alph{zahler}}
\renewcommand{\theequation}{\Alph{zahler}.\arabic{equation}}
\renewcommand{\thefigure}{\Alph{zahler}.\arabic{figure}}

\section{Allowed paths}\label{sec:choiceofpath}

Here, we discuss paths in the dispersive representation that allow for an undistorted path in the angular average, and follow the discussion in \cite{Aitchison:1966lpz}. For fixed $x$, the angular average amounts to a line integral 
between $h_\pm(x)$. Starting from $x=\infty$ to lower values, it is clear from Fig.~\ref{fig:2} that a critical value of $x$ is reached as soon as this straight line touches the value $4M_\pi^2$ for some $z\in [-1,1]$,
\begin{figure}[t]

   \begin{center}
~\hspace*{-0.7cm}    \includegraphics*[width=8.cm]{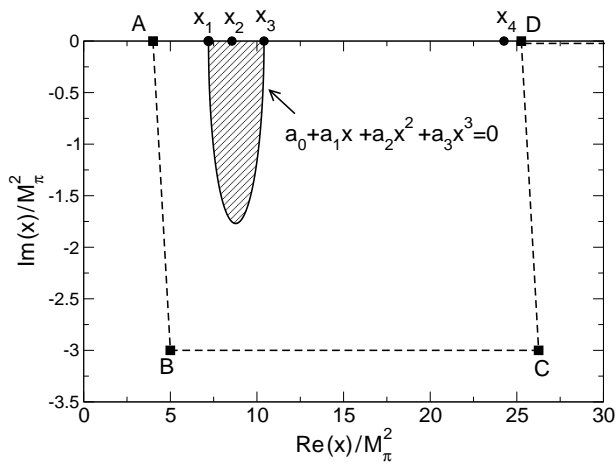}
  \end{center} 
\caption{
Solid line: The part of the complex  solutions of the polynomial equation (\ref{eq:Bpol}) with Im ($x$)$ < 0$.  Dashed: Deformed integration path used in the present work (displayed for 
$\lttwo{s}{30 M_\pi^2}$). Any deformed path that does not cross the solid line allows for an undistorted integration in the angular average (\ref{eq:angularaverages}).
 See appendix~\ref{sec:phaseshifts} for the actual values chosen for $A,\ldots,D$ and for $x_i$.
}
 \label{fig:B1}
 \end{figure}

\begin{figure}[t]
   \begin{center}
\hspace*{-0.7cm}    \includegraphics*[width=8.cm]{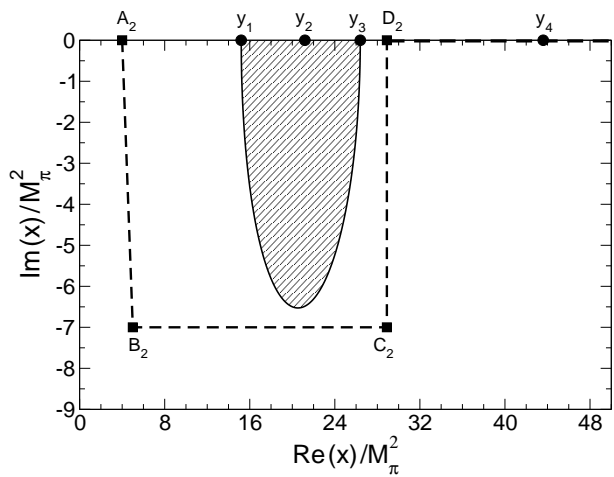}
  \end{center} 
\caption{The case $\omega\to 3\pi$. Solid line: Same as solid line in Fig~\ref{fig:B1},with $M_\eta^2$ replaced by $M_\omega^2$. Dashed: A deformed integration path. Any deformed path that does not cross the solid line allows for an undistorted integration in the angular average (\ref{eq:angularaverages}). For notation see appendix~\ref{sec:phaseshifts}. Note that $\sqrt{D_2} \simeq 0.75$ GeV, i.e., the Schenk parameterization is still operative at $D_2$ and generates  holomorphic phase shifts.} 
\label{fig:C2}
 \end{figure}

\bea
h(x,z)=4M_\pi^2\fs
\eea
This condition may be brought to the form
\bea\label{eq:Bpol}
\sum_{i=0}^3a_ix^i=0\scs\,\,\quad\quad -1\leq z \leq 1\scs
\eea
with
\bea
a_0&=&-4z^2M_\pi^2(M_\pi^2-M_\eta^2)^2\scs\nnnl
a_1&=&z^2(3M_\pi^2+M_\eta^2)^2-(M_\eta^2-5M_\pi^2)^2\scs\nnnl
a_2&=&-2z^2(M_\eta^2+3M_\pi^2)+2M_\eta^2-10 M_\pi^2\scs\nnnl
a_3&=&z^2-1\fs
\eea
For a fixed value of $z\neq \pm 1$, the equation (\ref{eq:Bpol}) has one real and 2 
complex conjugate solutions for $x$. Varying $z$ in the interval $(-1,1)$ traces out 3 curves in the complex $x$--plane. The real branch (Im ($x$)=0) is located on the negative real axis, Re($x)\leq 0$. In Fig.~\ref{fig:B1}, we display the part of the complex solutions with Im$(x) < 0 $ (solid black line).  If one uses a  path in the dispersion integral that does not cross the hatched region, the angular average can be performed in the undeformed interval $-1\leq z \leq 1$. The polygonal 
line ABCDEF (displayed for $s < 30 M_\pi^2$) is the version used in the present work. 
See appendix~\ref{sec:phaseshifts} for the actual values chosen for $A,\ldots,F$ and for $x_i$.

The following two remarks are in order at this place.

\begin{itemize}
\item[i)]
The reason to extend the polygonal line into the lower complex $x$--plane until $x=D$ is the following. After having solved the integral equation for the angular averages, one obtains the physical amplitudes by performing the final dispersive integration with Cauchy kernel $1/(x-s-i\epsilon)$. Even at $\epsilon=0$, this kernel is then not singular on the integration path for real values of $s$ with $s \not\in [D,F]$, except at the threshold
 $s=4M_\pi^2$. There, the integrand in (\ref{eq:M}) develops an integrable  square root singularity in case of  S-waves. It can be tamed with a transformation of variables, $z=A+(B-A)\tau^2\scs 0\le \tau\le 1$. The integration can thus be performed at $\epsilon=0$, without further ado -- an 
additional advantage of our procedure. For $s\in[D,F]$, due to the singularity of the Cauchy kernel, the angular  average needs to be interpolated between the pertinent Gauss--Legendre points, before the integration can be done. This is the reason why we push $D$ to higher values than required.  
\item[ii)]
In order to stay away from the threshold region $x=4M_\pi^2$ while performing the angular average, we have chosen the path BC with an imaginary part that is more negative than what is needed
according to the condition (\ref{eq:Bpol}), see also Figs.~\ref{fig:4} and \ref{fig:5}.
\end{itemize}

\setcounter{equation}{0}
\setcounter{section}{0}
\setcounter{figure}{0}
\addtocounter{zahler}{1}

\renewcommand{\thesection}{\Alph{zahler}}
\renewcommand{\theequation}{\Alph{zahler}.\arabic{equation}}
\renewcommand{\thefigure}{\Alph{zahler}.\arabic{figure}}

\section{The decay $\omega\to 3\pi$}\label{sec:omega}
We shortly consider the situation for $\omega \to 3\pi$ decays \cite{Niecknig:2012sj}. In Fig.~\ref{fig:C2}, we display the analogue
to the boundary in Fig.~\ref{fig:B1}, for $M_\eta^2$ replaced by $M_\omega^2$. A possibility for the integration path is displayed with  dashed lines. It is seen that with Schenk parameterization \cite{Schenk:1991xe}, it is still possible to avoid the pseudothreshold singularity and to perform the dispersive integral without interpolation between Gauss--Legendre points in the interval $s<D_2$, which contains the decay region. We therefore expect that the method also works in this case.         
\end{appendix}

\end{document}